\documentclass[useAMS,usenatbib]{mn2e}
\usepackage{graphicx}

\title[Dynamical analysis of the Blazhko effect]{Nonlinear dynamical analysis of the Blazhko effect with the \textit{Kepler} space telescope: the case of V783 Cyg}

\author[E. Plachy et al.]{E. Plachy$^{1,2}$\thanks{E-mail:
eplachy@astro.elte.hu}, J. M. Benk\H{o}$^{1}$,  Z. Koll\'ath$^{1,2}$, L. Moln\'ar$^{1,2}$ and R. Szab\'o$^{1}$\\
$^{1}$Konkoly Observatory, Research Centre for Astronomy and Earth Sciences, Hungarian Academy of Sciences,\\
 Konkoly Thege Mikl\'os \'ut 15-17, H-1121, Budapest, Hungary\\
$^{2}$Institute of Mathematics and Physics, Savaria Campus, University of West Hungary,\\ K\'arolyi G\'asp\'ar t\'er 4, H-9700 Szombathely, Hungary}
 
\begin{document}

\pagerange{\pageref{firstpage}--\pageref{lastpage}} \pubyear{2014}

\maketitle

\label{firstpage}

\begin{abstract}

We present a detailed nonlinear dynamical investigation of the Blazhko modulation of the \textit{Kepler} RR Lyrae star V783 Cyg (KIC 5559631). We used different techniques to produce modulation curves, including the determination of amplitude maxima, the O--C diagram and the analytical function method. We were able to fit the modulation curves with chaotic signals with the global flow reconstruction method. However, when we investigated the effects of instrumental and data processing artefacts, we found that the chaotic nature of the modulation can not be proved because of the technical problems of data stitching, detrending and sparse sampling. Moreover, we found that a considerable part of the detected cycle-to-cycle variation of the modulation may originate from these effects. According to our results, even the four-year-long, unprecedented \textit{Kepler} space photometry of V783 Cyg is too short for a reliable nonlinear dynamical analysis aiming at the detection of chaos from the Blazhko modulation. We estimate that two other stars could be suitable for similar analysis in the \textit{Kepler} sample and in the future TESS and PLATO may provide additional candidates.

\end{abstract}


\section{Introduction}

The Blazhko modulation has been one of the most puzzling mysteries in the field of stellar pulsation for more than a hundred years. A large fraction of known RR Lyrae stars show an inexplicable amplitude and phase variation. Some Cepheids show similar behaviour as well \citep{mk09,v473}. The origin of the modulation is still unclear. Ever since Sergey Blazhko discovered the phenomenon in RW Dra \citep{blazhko}, several explanations have been proposed, but none of them is without problems \citep{kovacs09,kolenberg12}. Studying the nature of the modulation may lead us to the right explanation. Some early models like the magnetic rotator model \citep{cousens,shibahashi00} and the nonradial resonant rotator model \citep{dziem,nowa} involve the rotation of the star and predict strictly periodic modulation. Recent observational evidences for irregular or multiperiodic modulation indicate that these models cannot explain the effect (see, e.g.\ \citealt{sodor11,gug12,benko14,skarka14}). Some models based on stochastically behaving phenomena, such as shockwaves \citep{gillet} or convection \citep{stothers}, suggest that the modulation must have a stochastic nature. On the other hand, resonant coupling between pulsation modes allows for chaotic behaviour, as it was shown by \citet{bk11}. Thus searching for chaos in modulation is a great opportunity to further constrain the theoretical models.    

Nonlinear dynamics became a remarkable research field since the phenomenon of chaos was initially reported \citep{lorenz}. The basic properties of chaotic behaviour are well-known: chaos may occur for certain parameters in deterministic systems that contain (even weak) nonlinearity. The sensitivity to the initial conditions makes these systems unpredictable. The chaotic behaviour appears as an irregular variation in the time series showing great similarity to stochastic ones, but the phase space geometry is very different. While a two dimensional cut of a phase space (a return map or Poincar\'e map) shows a random collection of points in the case of stochastic data, a chaotic map is always confined by a characteristically-shaped attractor that has fractal properties. Unfortunately, real observational data of irregular light variations are rarely long or precise enough to let its phase space pattern distinguish between chaotic, stochastic, or even multiperiodic nature. 

Despite of the difficulties, chaotic behaviour was already reported in several variable stars. Large-amplitude variations of late-type bright giant stars based on amateur observations collected by organisations such as the AAVSO (American Association of Variable Star Observers) and long-term photometric surveys provide data that are suitable for nonlinear investigations. Evidence of chaos was detected in some semiregular stars \citep*{semireg}, as well as in two RV Tauri-type \citep{RScuti,ACHer} and a Mira-type \citep{Kiss} variable star so far. Previously, chaos was not expected in weakly dissipative classical pulsators, based on theoretical predictions \citep{growth}. However, hydrodynamic models of W~Vir stars \citep{bk87,Serre2}, and the more recent calculations of RR Lyrae and BL Her models \citep*{plachy13,sm14} support that chaos can show up in classical variables as well, due to the nonlinear coupling of pulsation modes.

The detection of the routes to chaos also maintains a great interest in the field of nonlinear studies. There are different ways for chaos to evolve by varying a certain parameter of the system. The hydrodynamical models already produced period-doubling bifurcation, intermittency and crisis \citep{PDmodel, sm14}. In the period-doubling bifurcation scenario, chaos is preceded by a sudden change, where the new period is twice of the original period. Then the period bifurcates over and over again until the occurrence of the chaotic state. Intermittency manifests itself as a chaotic burst that irregularly interrupts a periodic behaviour. In crisis a transient chaos evolves to a permanently chaotic state. 
For more details about the properties of a simple chaotic systems we encourage the reader to study the summary presented by \citet{sm14}.  

The period-doubing bifurcation is detectable through the alternating maxima or minima in the light curve or through the subharmonic frequencies in the Fourier spectrum. It was observed in RV Tauri and white dwarf stars \citep*{preston63,goupil88}, and recently in RR Lyrae and BL Her stars as well \citep{pd,ujcikk,smolec12}. The former was discovered in the \textit{Kepler} prime mission and inspired new RR Lyrae model calculations as well as these investigations. Despite the numerical results from hydrodynamic models that had predicted chaos \citep{bk87,Serre2,plachy13,sm14}, chaotic behaviour in classical variable stars has not been reported yet. \textit{Kepler} may bring a new era to this field.

This study presents the first dynamical analysis of the Blazhko effect. Similar investigations have not been possible before, even with the persistent ground-based observations such as the Konkoly Blazhko Survey that were dedicated to this mysterious  phenomenon \citep{konkoly}, because of the lack of suitable data. Nonlinear investigations present serious requirements for the input data that can be hardly achieved from the ground. The requirements concern the continuity, the accuracy and the length. Quasicontinous data are suitable only if full pulsation cycles are covered and gaps do not affect the extrema significantly. This is rarely true for RR Lyrae stars, for which two consecutive cycles cannot be covered from most observing sites, except maybe from Antarctica \citep{antarctica}. Accuracy is especially important to distinguish chaotic data from periodic signals, whereas the length of the data is crucial to trace the divergence of the phase space trajectory. At least a few dozen modulation cycles are required to reliably detect chaos. However, stochastic data can resemble chaotic data on short lengths very easily. Any structure in the return maps of cycle-to-cycle variations can only be recognised if several points populate the map.

The possible irregular behaviour of Blazhko modulation is evident now. Photometric measurements of the CoRoT and \textit{Kepler} space telescopes contributed many additional details concerning the shape of the modulation \citep{benko14,szabo14,ujcikk}. The fact that period doubling and additional modes are detectable only in Blazhko stars points to a possible nonlinear mechanism. This directed our interest towards the nonlinear investigation of the modulation itself. Results obtained with the amplitude equation formalism already showed that the modulation can be chaotic \citep{bk11}. If chaos could be detected from observational data of the modulation, it would clearly support the idea of the mode-resonance model of the Blazhko effect. The photometry that \textit{Kepler} provides is the most suitable data for dynamical analysis so far. 

In this paper we focus on the star V783~Cyg (KIC~5559631). This star has the shortest amplitude-modulation period from the \textit{Kepler} sample, thus has the most modulation cycles observed through the four-year-long mission. Therefore it is the best candidate for our investigations. Our aim is to find out if any signs of chaos are detectable in the modulation of this star. In Section 2 we present the observed and the test data sets used in the analysis. The nonlinear analyser tool, the global flow reconstruction method is presented in Section 3. In Section 4 we show the analysis and summarise our results. Conclusions are drawn in Section 5.

\section{Data}

\begin{figure*}
 \includegraphics[width=178mm]{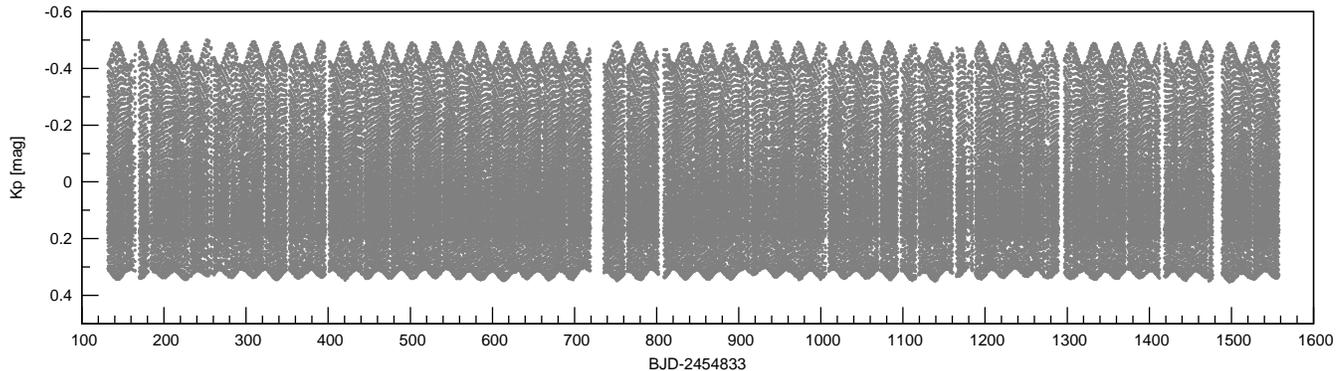}
 \caption{The final, tailor-made and rectified \textit{Kepler} light curve of V783 Cyg (KIC 5559631). }
\label{lc}
\end{figure*} 

V783 Cyg ($Kp=14.643$ magnitude, $\alpha_{2000}=19^h$ $52^m$ 52\fs74, $\delta_{2000}=+40\degr$ $47'$ 35\farcs4) is a fundamental-mode RR Lyrae star: an old, helium-burning, horizontal-branch star that is crossing through the classical instability strip. It has a pulsation period of $P= 0.6207001$ d and a modulation period of $P_m=27.67$ d. Although the observed Blazhko periods range from less than a week to a few years \citep{skarka13}, in the original \textit{Kepler} sample this star exhibits the shortest primary modulation cycle. Additional modes that might interfere with the variations of the fundamental radial mode were not detected by \citet{benko14}, down to $10^{-4}$ magnitudes in the Fourier spectrum in the star. 

The four-year-long prime mission of the \textit{Kepler} space telescope provided us with the most continuous photometric data of the RR Lyrae targets to date. Due to the unique precision, important new phenomena were discovered in Blazhko stars immediately after the first data release, such as period doubling and low-amplitude additional modes \citep{benko10,kolenberg10,pd}. Even with the unexpected end of the prime mission, unprecedented data were collected for a considerably long time almost continuously, allowing us to investigate the modulation in exquisite detail. To achieve the best photometric data a tailor-made aperture technique was developed by \citet{benko14}\footnote{http://konkoly.hu/KIK/data.html}. In this method all individual pixels were investigated around the star taking into account their contributions to the light curve. In our analysis we used the tailor-made light curve of V783 Cyg (KIC 5559631) that contains 16 quarters (Q1-Q16) of data observed in long cadence mode (30-minute sampling), presented in Figure \ref{lc}.

We performed the nonlinear dynamical analysis on the modulation itself instead of the actual light curve. Therefore the first step was the separation of the modulation from the pulsation. There are different methods to derive a modulation signal. Amplitude modulation can be easily determined from the amplitude maxima or minima of the cycles. The O--C diagrams of the same amplitude maxima and minima clearly show the phase modulation. The analytical function method provides an alternative technique by calculating time-dependent Fourier parameters \citep{gabor,kollath02}. We present four modulation curves that we found to be useful in the analysis and discuss their own uncertainties in more detail.

\subsection{Modulation curves from the light curve extrema}

\begin{figure}
 \includegraphics[width=84mm]{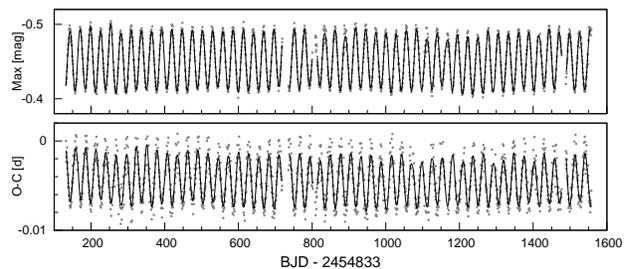}
 \caption{Top: variation of the amplitude maxima; bottom: variation in the O--C diagram. The solid lines are the spline fits used in the nonlinear analysis.}
\label{max-oc}
\end{figure}

We determined the amplitude maximum and minimum values of each pulsation cycle using a cubic spline technique. Due to the relatively sparse sampling ($\sim$~30~points/pulsation cycle) and missing data points, some fitted maxima contain significant error and had to be removed.

\subsubsection{Amplitude maxima} 
\label{amp_max}
The tailor-made photometry of \citet{benko14} processed all available pixels in the CCD mask of the stars, but there are indications that in some cases flux may have expanded outside the \textit{Kepler} target apertures. The flux loss of V783~Cyg was estimated to reach one percent, which can differ in the pulsation minima and maxima and vary from quarter to quarter. Much effort was taken to eliminate instrumental effects such as trends and amplitude scaling differences between quarters. These error sources may affect our dynamical analysis as they appear as cycle-to-cycle variations in the modulation curve. We found that since the modulation curve determined from the minima has smaller amplitudes, it is relatively more distorted by the instrumental effects than the one determined from the maxima. Therefore we chose the latter for the analysis (Figure \ref{max-oc}, upper panel).

\subsubsection{O--C diagram}

The phase values of the maxima, or equivalently, the \linebreak O--C values (Figure \ref{max-oc} lower panel), are expected to be less affected by the instrumental and data processing effects, but actually show a much larger scatter ($\sim$25\%) than the amplitude maximum values ($\sim$5\%).
Due to the 30-minute sampling, the pulsation cycles were strongly undersampled. This caused relatively high uncertainties in the maximum time determination. Therefore we first investigated if any cycle-to-cycle variation can be detected in the O--C diagram, which is a common feature among RR Lyrae stars \citep{gug12,benko14}.

We compared two fits to the O--C values: a periodic fit with a sine function and a spline fit that allows the cycle-to-cyle variation. We constructed the difference curve of the two fits and we found that its variation is smaller than the uncertainty of the O--C values ($\sim0.002$ d), so the detection of the cycle-to-cycle variation in the O--C diagram is not clear from the long cadence data alone.

\begin{figure}
 \includegraphics[width=84mm]{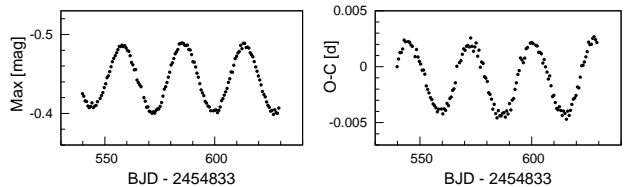}
 \caption{The amplitude maximum values (left) and the O--C diagram (right) of the short cadence data. A slight cycle-to-cycle variation is visible in both modulation curves.}
\label{scoc}
\end{figure}
   
Fortunately, V783 Cyg was observed in short cadence mode (1-minute sampling) in quarter 6 (Q6). This data set supports that the cycle-to-cycle variation in the O--C diagram exists (Figure \ref{scoc}). Therefore we decided to use the spline fit of the O--C diagram presented above as the second modulation curve to be analysed.

\subsection{Modulation curve derived with the analytical function method}

The analytical function method \citep{gabor} is a powerful tool to determine the time dependence of the amplitude and frequency of the pulsation modes. The analytical function is calculated in the Fourier space using a filtering window around the desired frequency \citep{kollath02}. 

\subsubsection{Analytical function of $R_{21}$}

When we calculated the amplitude variation curves with the analytical function method we realized that the instrumental effects discussed in Subsection \ref{amp_max} are able to fundamentally change the signal. After investigating the effects we found the stitching of the different quarters of the light curve to be the source of the largest distortions. We tested the problem by deriving a comparison light curve using a semi-automated $\chi^2$-minimization stitching technique that used linear scaling and zero-point shifts only. 

\begin{figure}
 \includegraphics[width=55mm,angle=270]{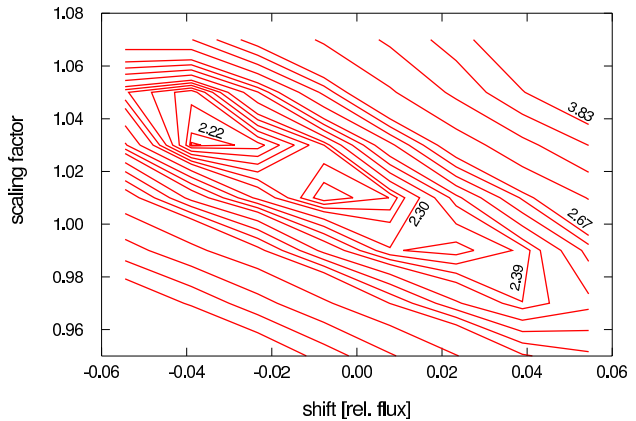}
 \caption{Contour plot of the $\chi^2$ values as a function of linear scaling and shifting of the light curve of a given \textit{Kepler} quarter designed to stitch it smoothly to the preceding one. Note the strong degeneracy between the parameters and the multiple minima with similar depths. Contours are labelled with  $\chi^2$ values.}
\label{chi}
\end{figure}

\begin{figure}
 \includegraphics[width=84mm]{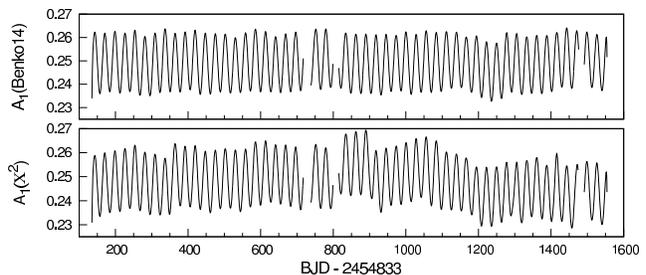}
 \caption{Comparison of the normalized-amplitude analytical functions of the light curves derived with the stitching technique of \citep{benko14} (upper panel) and our $\chi^2$ method (lower panel).}
\label{demon}
\end{figure}

We tried to fit two adjacent \textit{Kepler} quarters by adjusting two parameters. We fitted sine curves to the maxima and minima to one quarter of data, and applied the same functions to the next quarter by simply scaling and shifting the flux curve of the later quarter. We searched for the minimum of the $\chi^2$ values in this 2-dimensional parameter space. This is justified, since the envelopes  of the Blazhko modulation are reasonably sinusoidal, and there is no missing quarter between Q1 to Q16. In principle the method would provide a smooth stitching of adjacent quarters, but the strong degeneracy between the parameters hampered the determination of a unique solution except for a few quarter pairs. The linear relation between the two parameters suggested by the $\chi^2$ values (see Figure \ref{chi}) is useful, but the lowest minima do not always provide the best fit between the quarters, thus there is a remaining freedom to move along the trough. Fitting a more complex function to the envelopes may further constrain the fitting parameters, but that would require further (ad hoc) assumptions about the shape of the modulation from quarter to quarter which would complicate the procedure and is beyond the scope of this paper. \citet{benko14} employed a more empirical stitching method with moving-average-based detrending. Figure \ref{demon} displays the amplitude variations obtained from light curves derived by both methods. The \citet{benko14} method shows only a hint of patterns within quarters while the simple scale and shift method clearly shows such patterns. It turned out that linear scaling and shifts in themselves cannot remove all instrumental variations from the light curve.

However, this method does not distort the shapes of the pulsation cycles as does the detrending method applied by \citet{benko14}. We found that varying the moving average values also leads to differing distortions in the amplitude variation signal, that can reach 5-10\% differences in the modulation cycles. We note, however, that these are very small differences compared to the overall light variations ($\sim0.1\%$) and other studies, like a standard Fourier analysis, should be less affected by them. 

\begin{figure}
 \includegraphics[width=84mm]{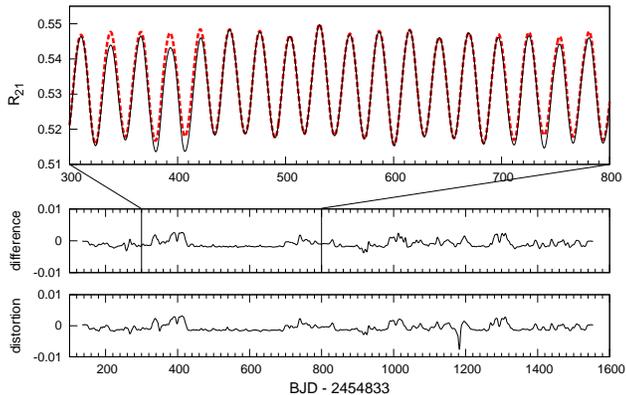}
 \caption{The upper panel shows a fraction of the $R_{21}=A_2/A_1$ modulation curve where the results of the method of \citet{benko14}, in black, and of our $\chi^2$ technique, in red (dashed), are identical (between time coordinates 420 and 700). Here the cycle-to-cycle variation can not be explained by instrumental effects, it must be real. The middle panel shows the difference of the red and black curves from the upper panel. The curve in the lower panel shows the distortion in the $R_{21}$ parameter caused by the detrending technique of \citet{benko14}.}
\label{instr}
\end{figure}

The distortion effects are minimized in the analytical function of ${R_{21}}$ which is determined as the quotient of the amplitude of the first harmonic and the main frequency, $R_{21}=A_2/A_1$ \citep{simonlee}. The differences between the two stitching methods in the ${R_{21}}$ curves come from the moving-average scaling, as we show in Figure \ref{instr}. We compared the difference curve of the methods to  an artificially created distortion curve. First, instrumental effects similar to the discontinuities in the raw \textit{Kepler} data were added to a periodically modulated signal that was then subsequently restored (see Subsection \ref{sect_test}). Then the difference between the $R_{21}$ curves before and after the corruption and restoration of the artificial signal were calculated. This provided us with the distortion curve of the method of \citet{benko14} (lower panel of Figure \ref{instr}). This curve is very similar to the difference curve in the middle panel of Figure \ref{instr}. This implies that the distortions are caused by the moving-average technique which is present only in the method of \citet{benko14}. Therefore the $\chi^2$-method version of the ${R_{21}}$ parameter was chosen as the third modulation curve for the analysis.

\subsubsection{Analytical function of the pulsation period ($P_1$)}
\label{p1}

We also calculated the temporal variation of the main pulsation period of the light curve with the analytical function method. We investigated the effects of noise on this modulation curve to estimate the uncertainties. We performed a Monte Carlo simulation of noisy, periodically modulated artificial data. 
We constructed this data from the Fourier-parameters of the main pulsation frequency of V783~Cyg  ($f_1=1.611$ c/d) and its harmonics, creating magnitudes at the times of \textit{Kepler} observations. We applied a periodic modulation by adding a sinusoidal amplitude modulation with the average amplitude of the observed one (0.0404~$Kp$ magnitudes). 
We then added different gaussian noise to the data with $\sigma=3$ mmag that corresponds to the residual noise after a 50-frequency fit to the observations. We note that the observational precision of V783 Cyg is one order of magnitude smaller ($\sigma=0.34$ mmag on the average) than the noise we used in the simulation.

We derived the analytical functions for every noisy artificial data sets and checked the cycle-to-cycle variation in the modulation. We defined the magnitude of the cycle-to-cycle variation by calculating the average difference of the maximum points between every second, third, fourth and fifth modulation cycles, respectively. We then compared the average differences of the observed modulation curve ($\sim 2\cdot10^{-5}$ d) to the ones we got from the Monte Carlo simulation ($\sim 7\cdot10^{-6}$ d), and found the former to be significantly larger. The latter value can be adopted as the uncertainty of the $P_1$ analytical function.

This investigation showed that beyond the uncertainties, a cycle-to-cycle variation exists in the analytical function of $P_1$. $P_1$ became the fourth modulation curve in the analysis.

\vskip 0.5 cm

\begin{figure}
 \includegraphics[width=84mm]{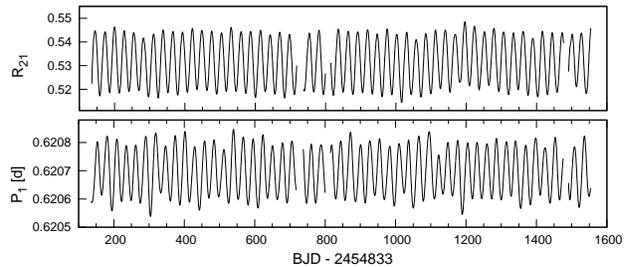}
 \caption{The variation of the $R_{21}$ Fourier amplitude relation, and the period variation of the light curve ($P_1$) derived by the analytical function method.}
\label{r21}
\end{figure}

We used the periodically modulated artificial data set to study also the data processing effects on the modulation curves.
Different trends and scaling factors, similar to the discontinuities and jumps in the observational data, were applied in every quarter to distort the periodically modulated data set. Then we used the stitching and detrending method of \citet{benko14} as an attempt to restore the original light curve. 

These investigations showed that instrumental effects distort not only the amplitude maxima and the ${R_{21}}$ parameter, but the period, $P_1$ too. Distortion manifests itself as additional cycle-to-cycle variation and its magnitude was estimated to be $6.8\cdot10^{-6}$ d due to the sparse sampling and  $1.5\cdot10^{-5}$ d due to the detrending method, respectively. This implies that a large part of the irregularity observed in the Blazhko modulation of V783 Cyg may have instrumental and data processing origins.

\subsection{Testing the instrumental and the data processing effects}
\label{sect_test}

\begin{figure}
 \includegraphics[width=84mm]{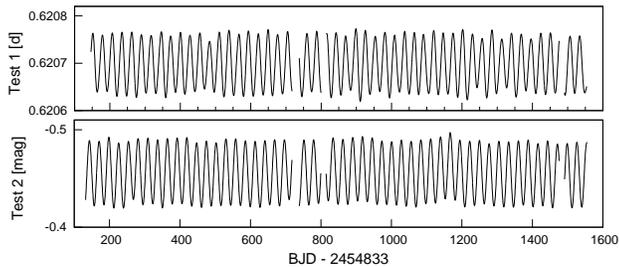}
 \caption{Test data sets. Top: the analytical function of the pulsation period of the periodically modulated test data set with $\sigma = 0.003$ gaussian noise added (\textit{Test 1}). Bottom: Spline-smoothed values of the pulsation cycle maxima after the periodically modulated test data set was distorted and then restored with the method of \citet{benko14} (\textit{Test 2}). }
\label{test}
\end{figure}

The irregularities arising from instrumental noise and data processing effects can mimic chaotic behaviour on short time scales. 
To check this possibility we chose two test curves for the nonlinear analysis along with the observed modulation data.  
We decided to test the effect of noise using a phase modulation curve, and the data processing effects with the amplitude modulation curve.
Because the modulation curve of the maxima has to be smoothed for the nonlinear analysis, it is less useful than an analytical function for testing the effects of noise. Therefore we used the analytical function of the period of a noisy, periodically modulated data set from the Monte Carlo simulation discussed in Section \ref{p1} (\textit{Test 1}). 
The other test data set was dedicated to test the stitching and detrending effects of \citet{benko14}. In this case we focused on the modulation curve of the amplitude maximum values (\textit{Test 2}). Test data sets are displayed in Figure \ref{test}.

\subsection{Return maps}

Return maps, i.e.\ Poincar\'e sections of phase space trajectories are very useful to visualize the differences between periodic, chaotic and stochastic data. We choose a characteristic point of the cycle and we plot the value of the succeeding point against the previous one. The maximum and minimum points of the amplitudes are ideal for this method. In a monoperiodic data set the values of the extrema do not change, they return to the same value, therefore the return map is single point. An irregular time series manifests itself as a set of points in the return map: the values of the extrema never repeat, so if we plot every extremum against the previous ones, the points will scatter in the map. According to the definition of chaos, the trajectories in the phase space have fractal structure that is also present in the return maps. However, the chaotic pattern is recognizable only if a large enough number of cycles populates the map. Otherwise it can mimic a random scatter which is typical for stochastic behaviour.             

\begin{figure}
 \includegraphics[width=84mm]{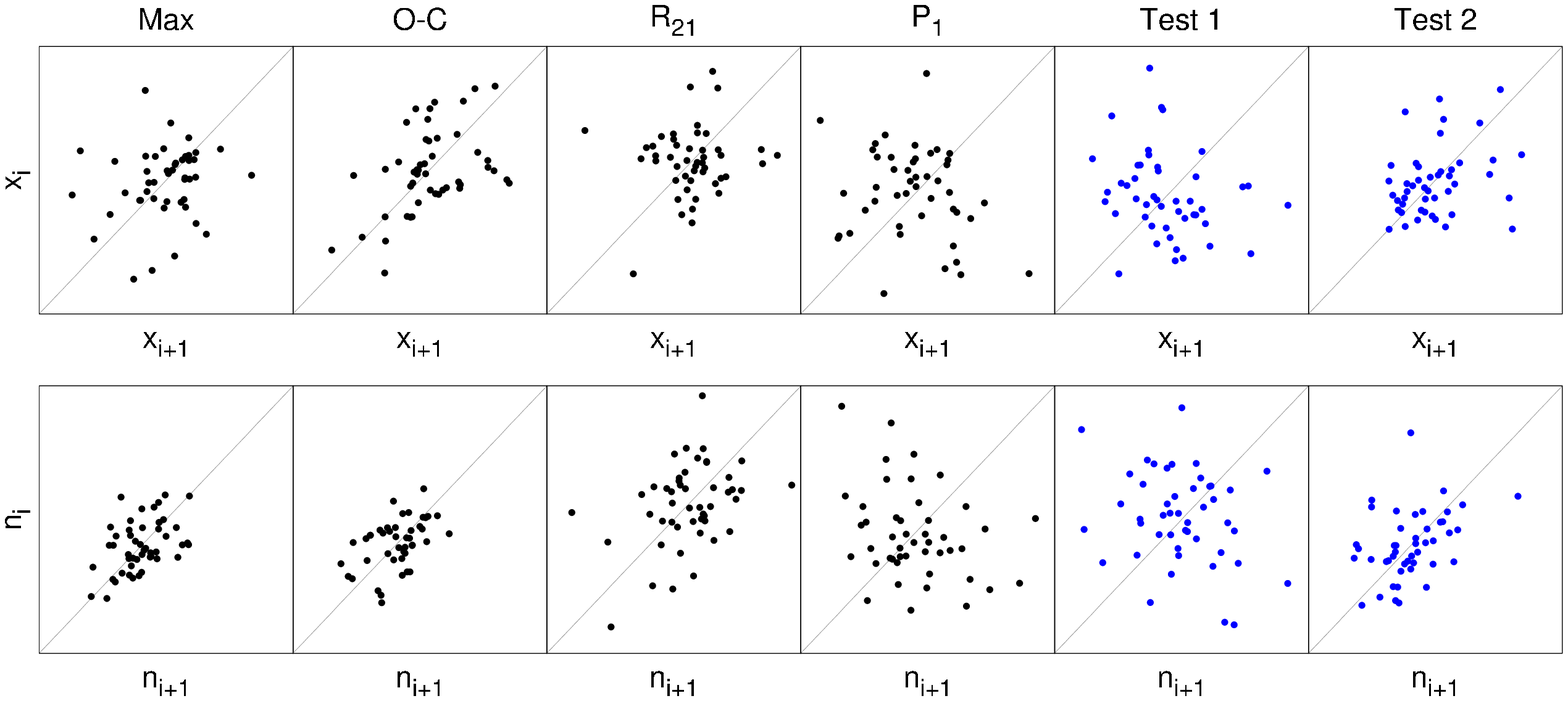}
 \caption{Return maps of the modulation curves (black) and the two test data sets (blue). Top row: maximum values, bottom row: minimum values. Given the small number of points, chaotic structure of the return maps cannot be recognized and distinguished from stochastic distributions.}
\label{returnmaps}
\end{figure}

We constructed the return maps of the four modulation curves (\textit{Max}, \textit{O--C}, $R_{21}$ and $P_1$) and the two test data sets (\textit{Test~1} and \textit{Test 2}) derived from the maximum and minimum values (Figure \ref{returnmaps}). The diagrams of the modulation curves do not show clear patterns, they practically can not be distinguished from the stochastic test data set. This suggests that the length of the data set i.e.\ the 51 modulation cycles are not satisfactorily long to construct an unambiguous return map. We note that the minimum length depends on both the amount of noise and the complexity of the system, therefore it is unclear how many cycles would be required in this case.    

\section{Method}

For the nonlinear analysis we used the global flow reconstruction method that was recently used in the study of the chaotic nature of RR Lyrae models as well \citep{plachy13}. This tool was already proved to be useful to detect low dimensional chaos not only in artificial but in real observed data too \citep{RScuti,semireg,ACHer}. For the detailed description of the method we refer to the original paper of \citet*{Serre}. Here we go through only the basic steps and configurations.

First we resample the data set with equal time spacing, ${s(t_n)}$, and produce the delay vectors, $\mathbf{X}(t_n)={(s(t_n), s(t_n-\Delta),s(t_n-2\Delta),...,s(t_n-(d_e-1)\Delta))}$, where $\Delta$ is the time delay and $d_e$ is the embedding dimension of the reconstruction space. Then we search for a map, $\mathbf{F}$, that connects the neighbouring points, evolving the trajectory in time: $\mathbf{X}^{n+1}=\mathbf{F}(\mathbf{X}^n)$. The existence of such map is a reasonable assumption. We use polinoms to calculate the map $\mathbf{F}$. After we found the map we are able to iterate synthetic data sets for arbitrary lengths. There is a reasonable length were the quantitative properties like the Lyapunov dimension can be determined with sufficient accuracy. The Lyapunov dimension characterizes the geometry of the chaotic attractor and refers to its complexity. It is calculated from the Lyapunov exponents that give the rate of divergence of phase space trajectories that were originally infinitesimally close to each other \citep{Ly}. At least one of the Lyapunov exponents must be positive according to the definition of chaos. Of course the quantitative properties of the synthetic data set of the iterated map can be adopted to the original data set only if great similarity is clearly visible between them.

The similarity is required in different visualizations simultaneously: we compare the synthetic and original data sets through the time series, the Fourier transforms and the orthogonal projections of the data sets following the Broomhead-King method \citep{BK} as well. At this point the method contains a subjective factor. For this reason, and to check the robustness of the reconstruction, we work in a large parameter space. We vary the time delay ($\Delta$) and the embedding dimension ($d_e$). We also add a small amount of gaussian noise to the data set and then we smooth it. This procedure stabilizes the reconstruction and provides two additional variable parameters: the noise intensity ($\xi$) and the spline smoothing parameter ($\sigma$). In this manner we can expand the search for the map in the close neighbourhood of the phase space trajectory. If we can identify a whole region of synthetic signals in the parameter space that resemble the original, we consider the reconstruction to be successful. The Lyapunov dimension of the original data set is most probably within the interval that we determined from the large number of synthetic signals.   

\begin{figure}
 \includegraphics[width=84mm]{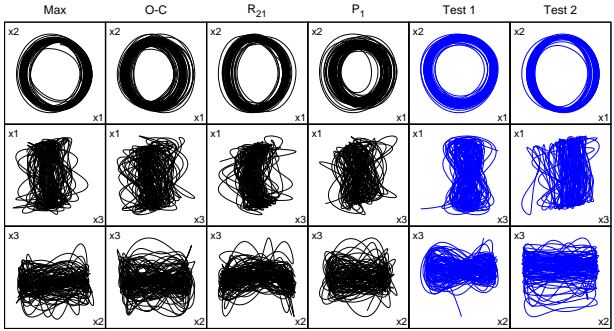}
 \caption{Broomhead-King projections of the four modulation curves and the two test data sets. All projections are very similar to each other.}
\label{BK}
\end{figure}

\section{Analysis}

We present the analysis of the four modulation curves (\textit{Max}, \textit{O--C}, $R_{21}$ and $P_1$) and the two test data sets (\textit{Test 1} and \textit{Test~2}). The Broomhead-King projections of the analysed data sets are displayed in Figure \ref{BK}. The parameters of the global flow reconstructions are given as follows: the time delay parameter, $\Delta = 4-30$, was fitted to the sampling of the data sets, the noise intensity was $\xi = 0-0.00009$ and the smoothing parameter was $\sigma = 0 - 0.009$. We applied all the embedding dimension values ($d_e$) that our method allows: 4, 5 and 6. This parameter space is relatively large, therefore we could obtain hundreds of chaotic maps. We used the same parameter settings for all of our modulation curves and test data. The curves were normalized and resampled equally to perform a homogeneous analysis. 

\begin{figure}
 \includegraphics[width=84mm]{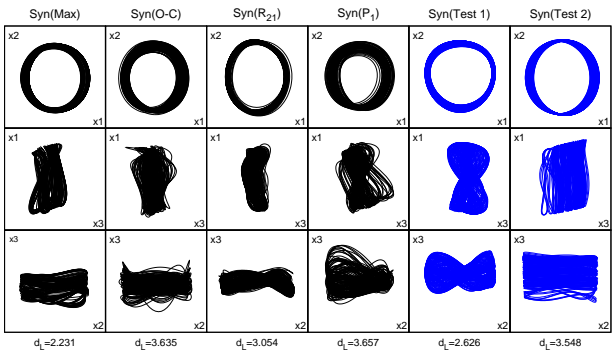}
 \caption{Broomhead-King projections of the best synthetic signals of the reconstruction of the four modulation curves and the two test data sets. The projections display a good similarity to the original ones in Figure \ref{BK}, and slight differences can be noticed between the various reconstructions. }
    \label{syn}
\end{figure}

\begin{table}
\caption{Results of the reconstructions. The columns indicate the numbers of chaotic maps obtained, the numbers and fractions of acceptable synthetic signals and their average Lyapunov dimensions for each input data set. }
\label{obs}
\footnotesize 
\begin{tabular}{l c c c c c}
\hline
& Chaotic maps & Acceptable & Fraction & $D_L$ \\
\hline
\textit{Max} & 154 & 143 & 93\% & 2.48$\pm$0.47 \\
 \textit{O--C} & 207 & 166 & 80\% & 2.63$\pm$0.54 \\
 $R_{21}$ & 647& 275 & 43\% & 2.43$\pm$0.37  \\
 $P_1$ & 500& 411 & 82\% & 2.46$\pm$0.44 \\
 \hline
 \textit{Test 1} & 301 & 227 & 75\% & 2.85$\pm$0.43 \\
 \textit{Test 2} & 166 & 139 & 84\% & 2.39$\pm$0.37 \\
\hline
\end{tabular}
\end{table}

We could fit all four modulation curves with chaotic signals. The best synthetic signals from each reconstructions are displayed in Figure \ref{syn}. We accepted those synthetic signals that resembled the original data sets and calculated their Lyapunov dimensions. Table \ref{obs} summarises our findings. (The results of the reconstructions with different embedding dimensions were merged.) We note that the precise values of the number of synthetic signals are not relevant, these numbers can be changed by using different initial conditions. It is the magnitude  and the fraction of acceptable synthetic signals that indicate the robustness of the reconstruction. 

We could also fit both test data sets with chaotic signals. This implies that a periodic data set distorted by noise or instrumental effects, and which has the same length and sampling as the observational data set, can be deceptively similar to a chaotic signal.

We calculated the average Lyapunov dimension values from the accepted synthetic signals and we used the standard deviation values as uncertainties. However, the quantitative properties of the chaotic behaviour in the modulation of this star could be determined with a large scatter only. This is not surprising from a relatively short data set. It is undefinable whether the Lyapunov dimension is between 2--3 or 3--4.  We tried to constrain the result further by determining the minimum embedding dimension with an alternative method. 

\begin{figure}
 \includegraphics[width=84mm]{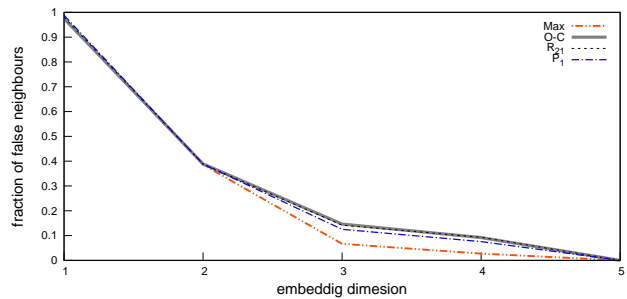}
 \caption{Estimation of the minimal embedding dimension by the the false nearest neighbour algorithm. The results of the four modulation curves are displayed with different colors. The value of the embedding dimension is very uncertain.}
\label{fnn}
\end{figure}

The minimum embedding dimension, that must be larger than the Lyapunov dimension, can be easily estimated by the false nearest neighbour algorithm proposed by \citet{Kennel}. The number of neighbours of a point along the trajectory decreases with increasing embedding dimension. If the embedding dimension is lower than necessary, the trajectories in the phase space appear entangled and many of the neighbouring points will be false neighbours, but in an appropriate dimension (or higher), they will be separated properly. We plotted the calculated fractions of false nearest neighbours for each modulation curve in Figure \ref{fnn}. For noiseless data the fraction of false neighbours must decrease dramatically, approaching zero at the appropriate dimension, but the picture is not so clear in our case. The minimum embedding dimension of the modulation cannot be determined unambiguously, it can be anywhere from $d_e > 2$ to $d_e=5$. Therefore we cannot constrain the Lyapunov dimension with this method either.

\section{Conclusions}

We investigated the modulation of the Blazhko RR Lyrae star V783 Cyg (KIC 5559631), based on four years of observations from the \textit{Kepler} space telescope. We found that the determination of modulation curves is not unambiguous, several problems affect the nonlinear analysis:

\begin{itemize}
\item The flux loss due to the tight pixel aperture has an unknown effect on amplitudes.
\item The amplitudes are highly sensitive to quarter stitching techniques: scaling and shifting cause significant distortions, especially if a moving average is used.
\item The instrumental and data processing distortions affect the pulsation maximum values, but alters the analytical functions of the Fourier amplitudes even more so.
\item The $R_{21}$ amplitude ratio of the first harmonic and the main pulsation frequency is also very sensitive to the moving-average technique.
\item Due to the sparse sampling of long cadence data, the O--C diagram can be determined with large scatter only, and even the cycle-to-cycle variation itself cannot be clearly detected.
\item The noise and the data processing effects affect the analytical function of the main pulsation period as well.
\end{itemize} 

We studied the effect of all instrumental problems and we concluded that even if they cause significant distortions, a cycle-to-cycle variation definitely exists in the Blazhko modulation. Our aim was to define whether these irregularities are stochastic or chaotic. We analysed four modulation curves using the global flow reconstruction method: the variation of the maxima, the O--C diagram, the $R_{21}$ amplitude relation curve, and the analytical function (temporal variation) of the main period. We could fit all these modulation curves with chaotic signals. The Lyapunov dimension of the signals spread out to a wide range: 2.001-3.635. The data set is too short to determine the quantitative properties of the chaotic dynamics with higher precision. 

On the other hand, we constructed test signals to study the reliability of the results of the nonlinear analysis. The effects of stochastic noise and data processing effects were included. We found that the cycle-to-cycle variation caused by these effects can be also fitted by chaotic signals similar to the modulation curves. This clearly shows that our results from the nonlinear dynamical analysis of V783 Cyg are very uncertain. Even the unprecedented \textit{Kepler} data of this star is not precise and long enough to analyse the irregular nature of the modulation with nonlinear methods. Thus in this case we cannot confirm that the underlying dynamics of the modulation is governed by chaotic processes.  

We have now started to explore whether more potential Blazhko targets for nonlinear analysis exist. Other stars in the \textit{Kepler} sample have higher modulation amplitudes and therefore provide less noisy data for the analysis, but at the cost of less modulation cycles covered. Based on our estimates only one or two more stars are potentially suitable for a similar analysis.

The coverage, length and precision of the \textit{Kepler} measurements will be unmatched for years to come. Ground-based observations produce inferior data: in the best case, several more years of observations will be required before a similar nonlinear analysis could be attempted. Long-duration, quasi-continuous observations will be produced only in the Continuous Viewing Zones of the TESS space telescope \citep{tess} and later by the PLATO space telescope \citep{plato}, allowing for a renewed effort to detect chaos in the Blazhko effect a few years from now. 

\section*{Acknowledgements}

We thank the referee for her/his useful comments. 
This work has been supported by the Hungarian OTKA grant K83790.
The work of E. Plachy leading to this research was supported by the European Union and the State of Hungary, co-financed by the European Social Fund in the framework of T\'AMOP 4.2.4.\ A/2-11-1-2012-0001 `National Excellence Program'. The research leading to these results has received funding from the European Community's Seventh Framework Programme (FP7/2007-2013) under grant agreements no. 269194 (IRSES/ASK) and no. 312844 (SPACEINN) and from the ESA PECS Contract No. 4000110889/14/NL/NDe. R. Szab\'o\ acknowledges the J\'anos Bolyai Scholarship of the Hungarian Academy of Sciences.


\begin{thebibliography}{}

\bibitem[\protect\citeauthoryear{Abarbanel et al.}{1993}]{Ly} Abarbanel H. D. I., Brown R., Sidorowich J. J., Tsimring, L. S., 1993, RevModPhys, 65, 1331 
\bibitem[\protect\citeauthoryear{Benk\H{o} et al.}{2010}]{benko10} Benk\H{o} J. M. et al,  2010, MNRAS, 409, 1585
\bibitem[\protect\citeauthoryear{Benk\H{o} et al.}{2014}]{benko14} Benk\H{o} J. M., Plachy E., Szab\'o R., Moln\'ar L., Koll\'ath Z.,  2014, ApJS, 213, 31
\bibitem[\protect\citeauthoryear{Blazhko}{1907}]{blazhko} Blazhko S., 1907, AN, 175, 325
\bibitem[\protect\citeauthoryear{Broomhead \& King}{1987}]{BK}Broomhead D. S., King G. P., 1987, Physica D, 20, 217
\bibitem[\protect\citeauthoryear{Buchler \& Kov\'acs}{1987}]{bk87} Buchler J. R., Kov\'acs G., 1987, ApJ, 320, 57
\bibitem[\protect\citeauthoryear{Buchler \& Koll\'ath}{2011}]{bk11} Buchler J. R., Koll\'ath Z., 2011, ApJ, 731, 24
\bibitem[\protect\citeauthoryear{Buchler, Koll\'ath \& Cadmus}{Buchler et al.}{2004}]{semireg} Buchler J. R., Koll\'ath Z., Cadmus R. R. Jr., 2004, ApJ, 613, 532
\bibitem[\protect\citeauthoryear{Buchler et al.}{1996}]{RScuti} Buchler J. R., Koll\'ath Z., Serre T., Mattei J., 1996, ApJ, 462, 489
\bibitem[\protect\citeauthoryear{Cousens}{1983}]{cousens} Cousens A., 1983,
MNRAS, 203, 1171
\bibitem[\protect\citeauthoryear{Dziembowski \& Cassisi}{1999}]{dziem} Dziembowski W., Cassisi S., 1999, AcA, 49, 371
\bibitem[\protect\citeauthoryear{G\'abor}{1946}]{gabor} G\'abor D., 1946, J. Inst. Electr. Eng. III, 93, 429
\bibitem[\protect\citeauthoryear{Gillet}{2013}]{gillet} Gillet D., 2013, A\&A, 554, 46
\bibitem[\protect\citeauthoryear{Goupil, Auvergne \& Baglin}{Goupil et al.}{1988}]{goupil88} Goupil M. J., Auvergne M., Baglin A., 1988, A\&A, 196, 13
\bibitem[\protect\citeauthoryear{Guggenberger et al.}{2012}]{gug12} Guggenberger E. et al., 2012, MNRAS, 424, 649
\bibitem[\protect\citeauthoryear{Jurcsik et al.}{2009}]{konkoly} Jurcsik J., et al., 2009, MNRAS, 400, 1006
\bibitem[\protect\citeauthoryear{Kennel, Brown \& Abarbanel}{Kennel et al.}{1992}]{Kennel} Kennel M., Brown R., Abarbanel H., 1992, Physical Review A, 45, 3403
\bibitem[\protect\citeauthoryear{Kiss \&  Szatm\'ary}{2002}]{Kiss} Kiss L. L., Szatm\'ary, K., 2002, A\&A, 390, 585
\bibitem[\protect\citeauthoryear{Kolenberg}{2012}]{kolenberg12} Kolenberg K., 2012, JAAVSO, 40, 481
\bibitem[\protect\citeauthoryear{Kolenberg et al.}{2010}]{kolenberg10} Kolenberg K. et al., 2010, ApJ, 713, L198
\bibitem[\protect\citeauthoryear{Koll\'ath, Moln\'ar \& Szab\'o}{Koll\'ath et al.}{2011}]{PDmodel} Koll\'ath Z., Moln\'ar L., Szab\'o R., 2011, MNRAS, 414, 1111
\bibitem[\protect\citeauthoryear{Koll\'ath et al.}{1998}]{ACHer} Koll\'ath Z., Buchler J. R., Serre T., Mattei J., 1998, A\&A, 329, 147
\bibitem[\protect\citeauthoryear{Koll\'ath et al.}{2002}]{kollath02} Koll\'ath Z., Buchler J. R., Szab\'o R., Csubry Z., 2002, A\&A, 385, 932
\bibitem[\protect\citeauthoryear{Kov\'acs}{2009}]{kovacs09} Kov\'acs G., 2009, AIPC, 1170, 261 
\bibitem[\protect\citeauthoryear{Lorenz}{1963}]{lorenz} Lorenz E. N., 1963, Journ. of Atm. Sci., 20, 130
\bibitem[\protect\citeauthoryear{Moln\'ar \& Szabados}{2014}]{v473} Moln\'ar L., Szabados, L. 2014, MNRAS, 422, 3222
\bibitem[\protect\citeauthoryear{Moskalik \& Ko{\l}aczkowski}{2009}]{mk09} Moskalik P., Ko{\l}aczkowski Z. 2009, MNRAS, 394, 1649
\bibitem[\protect\citeauthoryear{Nowakowski \& Dziembowski}{2001}]{nowa} Nowakowski R. M., Dziembowski W. A., 2001, AcA, 51, 5
\bibitem[\protect\citeauthoryear{Plachy, Koll\'ath \& Moln\'ar}{Plachy et al.}{2013}]{plachy13} Plachy E., Koll\'ath Z., Moln\'ar L., 2013, MNRAS, 433, 3590
\bibitem[\protect\citeauthoryear{Preston et al.}{1963}]{preston63}Preston G. W., Krzeminski W., Smak J. Williams J. A., 1963, ApJ, 137, 401 
\bibitem[\protect\citeauthoryear{Rauer et al.}{2014}]{plato} Rauer H., et al., 2014, Exp. Astr., accepted, arXiv:1310.0696
\bibitem[\protect\citeauthoryear{Ricker et al.}{2014}]{tess} Ricker G. R., et al., 2014, Proc.\ SPIE, submitted, arXiv:1406.0151
\bibitem[\protect\citeauthoryear{Shibahashi}{2000}]{shibahashi00} Shibahashi, H., 2000, ASPC, 203, 299
\bibitem[\protect\citeauthoryear{Serre, Koll\'ath \& Buchler}{Serre et al.}{1996a}]{Serre2} Serre T., Koll\'ath Z., Buchler J. R., 1996a, A\&A, 311, 845
\bibitem[\protect\citeauthoryear{Serre, Koll\'ath \& Buchler}{Serre et al.}{1996b}]{Serre} Serre T., Koll\'ath Z., Buchler J. R., 1996b, A\&A, 311, 833
\bibitem[\protect\citeauthoryear{Simon \& Lee}{1982}]{simonlee} Simon N. R., Lee A. S., 1982, ApJ, 248, 291
\bibitem[\protect\citeauthoryear{Skarka}{2013}]{skarka13} Skarka, M., 2013, A\&A, 549, 101
\bibitem[\protect\citeauthoryear{Skarka}{2014}]{skarka14} Skarka, M., 2014, A\&A, 562, 90
\bibitem[\protect\citeauthoryear{Smolec \& Moskalik}{2014}]{sm14} Smolec R., Moskalik P., 2014, MNRAS, 441, 101
\bibitem[\protect\citeauthoryear{Smolec et al.}{2012}]{smolec12} Smolec R., et al., 2012, MNRAS, 419, 2407
\bibitem[\protect\citeauthoryear{S\'odor et al.}{2011}]{sodor11} S\'odor \'A. et al., 2011, MNRAS, 411, 1585
\bibitem[\protect\citeauthoryear{Stellingwerf}{1984}]{growth} Stellingwerf, R. F., 1984, ApJ, 277, 322
\bibitem[\protect\citeauthoryear{Stothers}{2006}]{stothers} Stothers R. B., 2006, ApJ, 652, 643
\bibitem[\protect\citeauthoryear{Szab\'o et al.}{2010}]{pd} Szab\'o, R. et al., 2010,
MNRAS, 409, 1244
\bibitem[\protect\citeauthoryear{Szab\'o}{2014}]{szabo14} Szab\'o, R. 2014, IAUS, 301, 241
\bibitem[\protect\citeauthoryear{Szab\'o et al.}{2014}]{ujcikk} Szab\'o, R. et al., 2014, A\&A, accepted
\bibitem[\protect\citeauthoryear{Wang et al.}{2011}]{antarctica} Wang L., et al., 2011, AJ, 142, 155
\end{thebibliography}
\end{document}